\documentclass{jpsj3}
\usepackage{txfonts}

\usepackage{bm}
\usepackage{multirow}

\newcommand{\fnm}{\footnotemark}

\newcommand{\scrs}{\scriptsize}

\newcommand{\ph}{{\rm phonon}}
\newcommand{\kph}{\kappa_\ph}

\newcommand{\vph}{v_\ph}

\newcommand{\tph}{\tau_\ph}

\newcommand{\spin}{{\rm spin}}
\newcommand{\ksp}{\kappa_\spin}

\newcommand{\Csp}{C_\spin}
\newcommand{\vsp}{v_\spin}

\newcommand{\kc}{\kappa_{\parallel c}}
\newcommand{\kp}{\kappa_{\perp c}}

\newcommand{\im}{{\rm i}}

\newcommand{\Boltz}{k_{\rm B}}
\newcommand{\Debye}{\varTheta_{\rm D}}

\newcommand{\TNone}{T_{\rm N1}}
\newcommand{\TNtwo}{T_{\rm N2}}
\newcommand{\Tsone}{T_{\rm s1}}
\newcommand{\Tstwo}{T_{\rm s2}}

\newcommand{\D}{{\rm d}}
\newcommand{\Ham}{{\mathcal H}}
\newcommand{\sx}{s^{x}}
\newcommand{\sy}{s^{y}}
\newcommand{\sz}{s^{z}}

\newcommand{\SCO}{Sr$_2$CuO$_3$}

\newcommand{\SrCO}{SrCuO$_2$}

\newcommand{\LaCuO}{La$_2$CuO$_4$}

\newcommand{\BaCoVO}{BaCo$_2$V$_2$O$_8$}

\newcommand{\ACoX}{$A$Co$X_3$}
\newcommand{\RCoC}{RbCoCl$_3$}
\newcommand{\RCoB}{RbCoBr$_3$}
\newcommand{\CCoC}{CsCoCl$_3$}
\newcommand{\CCoB}{CsCoBr$_3$}

\newcommand{\AXset}{($A$~=~Rb,~Cs; $X$~=~Cl,~Br)}

\newcommand{\onlinecite}[1]{\hspace{-1 ex} \nocite{#1}\citenum{#1}} 

\renewcommand{\surname}[1]{#1}

\title{
	Observation of the Thermal Conductivity due to Spins
	in the One-Dimensional Antiferromagnetic Ising-Like Spin System \\
	{\ACoX} {\AXset}
}

\author{
\name{Yoshiharu \surname{Matsuoka}}$^1$,
\name{Takayuki \surname{Kawamata}}$^1$\thanks{tkawamata@teion.apph.tohoku.ac.jp},
\name{Koki \surname{Naruse}}$^1$,
\name{Masumi \surname{Ohno}}$^1$,
\name{Yoichi \surname{Nishiwaki}}$^2$,
\name{Tetsuya \surname{Kato}}$^3$,
\name{Takahiko \surname{Sasaki}}$^4$,
\name{Yoji \surname{Koike}}$^1$
}
\inst{
	$^1$Department of Applied Physics, Tohoku University,
	6-6-05 Aoba, Aramaki, Aoba-ku, Sendai 980-8579, Japan \\
	$^2$Department of Physics, Saitama Medical University,
	Moroyama, Saitama 350-0495, Japan \\
	$^3$Faculty of Education, Chiba University,
	Yayoi-cho 1-33, Inageku, Chiba 263-8522, Japan \\
	$^4$Institute for Materials Research, Tohoku University,
	2-1-1 Katahira, Aoba-ku, Sendai 980-8577, Japan
}

\abst{
	Thermal conductivity measurements have been carried out for single crystals
	of the one-dimensional (1D) antiferromagnetic (AF) Ising-like spin system
	{\ACoX} {\AXset}.
	A shoulder originating from the thermal conductivity due to spin excitations,
	$\ksp$, has been observed around 60~K in the temperature dependence
	of the thermal conductivity along the $c$-axis parallel to spin chains
	for every {\ACoX}.
	To our knowledge,
	this is the first observation of $\ksp$ in 1D AF Ising-like spin systems.
	It has been found that
	the magnitude of $\ksp$ in {\ACoX} is a little smaller than
	those in 1D AF Heisenberg spin systems.
	The reason is discussed in terms of the Heisenberg-like character in {\ACoX}.
}

\begin{document}
\maketitle

\newpage

\section{Introduction}

The thermal conductivity in low-dimensional quantum spin systems
has attracted great interest,
because various remarkable anomalies associated with
the low dimensionality of the spin correlation have been observed
in these spin systems.
In particular, the thermal conductivity due to spin excitations, $\ksp$,
has been observed markedly in one-dimensional (1D) antiferromagnetic (AF)
Heisenberg spin systems,
\cite{Miike1975a,Kudo1999,Sologubenko2000,Sologubenko2000b,Sologubenko2000a,
Sologubenko2001,Kudo2001,Kudo2001a,Hess2001,Sologubenko2003a,
Hess2004a,Hess2006,Kawamata2008,Kawamata2010,N.Hlubek2010,Hlubek2010,Hlubek2012}
two-dimensional AF Heisenberg spin systems
\cite{Nakamura1991,Takenaka1997,Sun2003,Yan2003,
Hess2003a,Hess2003,Hofmann2003,Hess2004,Berggold2006}
and a 1D ferromagnetic Heisenberg spin system also.
\cite{Kudo2004,Choi2004}
It has been found that the magnitude of $\ksp$ tends to increase
with increasing exchange interaction between the nearest neighboring spins
in 1D AF Heisenberg spin systems.

The Hamiltonian of a 1D AF system is described using the $XXZ$ Heisenberg model
as follows:
\begin{equation}
	\Ham = J \sum_{i}^{\rm chain} \left[ \sz_{i}\sz_{i+1}
		 + \varepsilon
		   \left( \sx_{i}\sx_{i+1} + \sy_{i}\sy_{i+1} \right) \right],
	\label{eq:H}
\end{equation}
where ${\bm s} = (\sx, \sy, \sz)$ is the spin,
$J$ the intrachain exchange interaction between the nearest neiboring spins
and $\varepsilon$ the anisotropy parameter of the spin.
When spins are isotropic, namely, $\varepsilon = 1$,
the spin system is called a Heisenberg one.
When $\varepsilon = 0$, on the other hand,
the spin system is called an Ising one.
Using the spin raising and lowering operators,
$s^+ = \sx + \im\sy$ and $s^- = \sx - \im\sy$, respectively,
Eq.~(\ref{eq:H}) is rewritten as follows:
\begin{equation}
	\Ham = J \sum_{i}^{\rm chain} \left[ \sz_{i}\sz_{i+1}
		 + \frac{\varepsilon}{2}
		   \left( s^+_{i} s^-_{i+1} + s^-_{i} s^+_{i+1} \right) \right].
	\label{eq:H2}
\end{equation}
The second and third terms are so related to the propagation of spin excitations
as to be likely important for the appearance of $\ksp$.
Therefore, $\ksp$ is supposed to be very small in Ising spin systems.
Theoretically, in fact, this is supported.\cite{Huber1969}
Experimentally, no $\ksp$ has been observed
in the quasi-1D AF spin system {\BaCoVO}
where $\varepsilon \sim 0.5$. \cite{Zhao2012,Kolland2013}

The compound {\ACoX} {\AXset} is regarded as
a quasi-1D AF Ising-like spin system
where the space group of the crystal structure is $P6_3/mmc$ at room temperature.
The AF spin chains consist of face-shared Co$X_6$ octahedra
stacked along the $c$-axis,
and the spin chains compose a triangular lattice in the $c$-plane.
The Ising-like anisotropy of Co$^{2+}$ spins along the $c$-axis is
due to the spin-orbit interaction.
The Hamiltonian of {\ACoX} is given by Eq.~(\ref{eq:H}),
where $\varepsilon$ is estimated from
Raman scattering, \cite{Lehmann1981,Lockwood1983,Matsubara1991a}
neutron scattering, \cite{Yoshizawa1981,Nagler1983,Matsubara1991}
magnetization \cite{Hori1990}
and ESR measurements \cite{Shiba2003}
to be below 0.24 and smaller than that of {\BaCoVO},
as listed in Table~\ref{tbl:ACoX-param}.
It is known that {\ACoX} exhibits successive phase transitions
at low temperatures $\TNone$ and $\TNtwo$,
owing to the interchain interaction.
\cite{Lockwood1983,Yoshizawa1979,Yelon1975,Nishiwaki2008}
The magnetic state is paramagnetic at high temperatures above $\TNone$.
At intermediate temperatures between $\TNone$ and $\TNtwo$,
a partially ordered phase is formed,
where two thirds of spin chains are ordered like a honeycomb lattice in the $c$-plane
owing to the AF interchain interaction.
At low temperatures below $\TNtwo$,
all Co$^{2+}$ spins are ordered, so that a ferrimagnetic order is formed.
Moreover, it is known that {\RCoB} undergoes two structural transitions
at $\Tsone$ and $\Tstwo$. \cite{Yamanaka2002a}
Values of $\TNone$, $\TNtwo$, $\Tsone$ and $\Tstwo$ are
listed in Table~\ref{tbl:ACoX-param}
together with those of $J$ and the ratio of the interchain interaction, $J'$, to $J$.

In this paper,
in order to investigate
the contribution of spin excitations
to the thermal conductivity in 1D AF Ising-like spin systems,
we have measured the thermal conductivity of {\ACoX} {\AXset}.

\section{Experimental}

Single crystals of {\ACoX} {\AXset} were grown by the Bridgman technique
using equimolar mixture of {$AX$} and Co{$X_2$}
sealed in an evacuated quartz tube.
Both of the thermal conductivity along the $c$-axis parallel to spin chains, $\kc$,
and that in the $c$-plane perpendicular to spin chains, $\kp$,
were measured by the conventional steady-state method.
The typical dimensions of a single crystal used for the measurements were
4~mm along the direction of heat current and
2$\times$2~mm{$^2$} perpendicular to the direction.
One side of a single crystal was anchored
on the copper heat sink with indium solder.
A chip-resistance of 1~k$\Omega$ (Alpha Electronics MP1K000) was
attached as a heater to the opposite side of the single crystal
with Araldite epoxy adhesive.
The temperature gradient across the crystal was
measured with two Cernox thermometers (Lake Shore Cryotronics, CX-1050-SD).
The error of the absolute value of the thermal conductivity
obtained was estimated to be about 10~\%
on account of errors of the crystal geometry.
Thermal conductivity measurements in magnetic fields up to 14~T were
also carried out for {\RCoC} using a superconducting magnet.

\section{Results and Discussion}

\begin{figure}
	\includegraphics[width=0.95\textwidth]{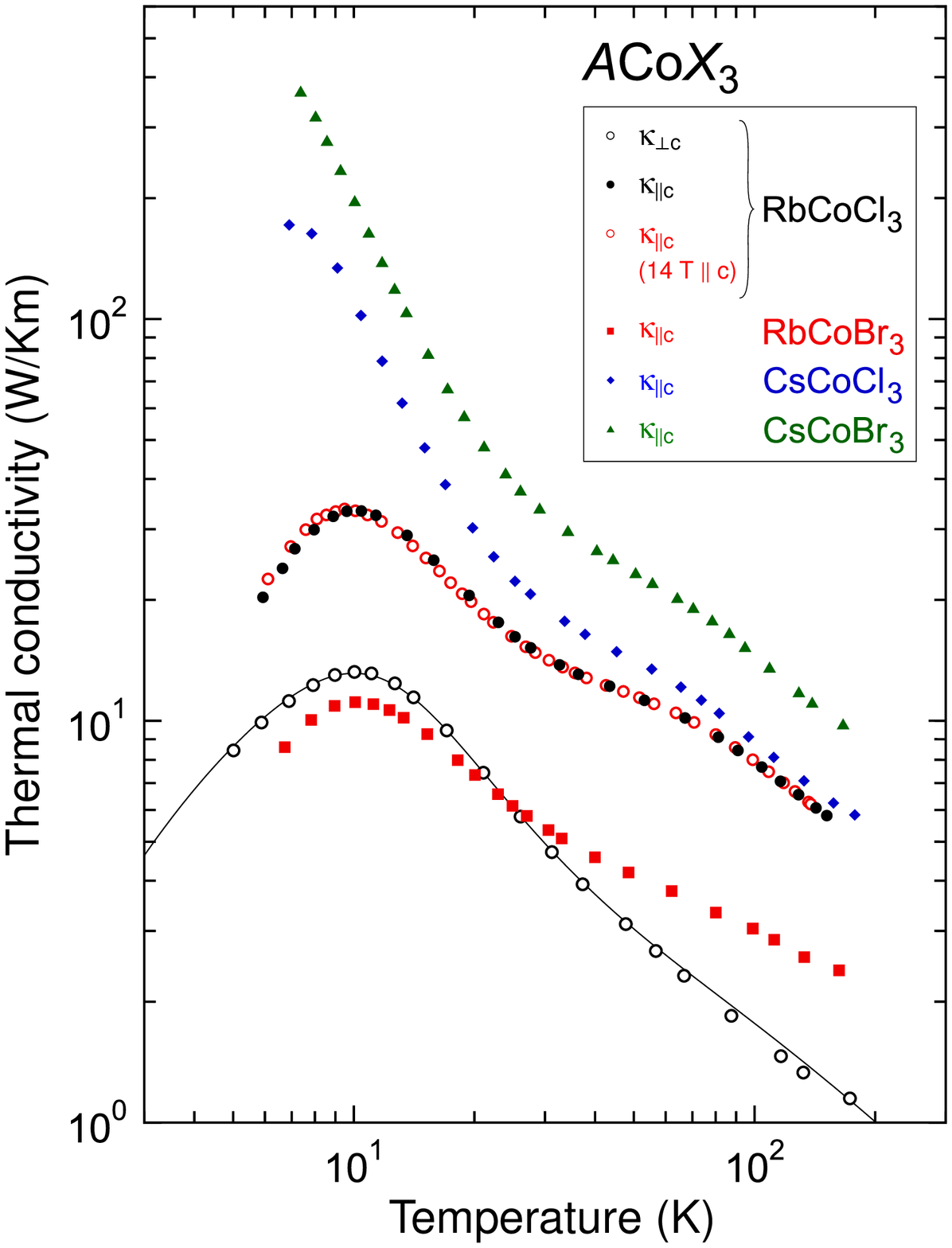}
	\caption{}\label{fig:ACoX-kappa}
\end{figure}

Figure~\ref{fig:ACoX-kappa} shows the temperature dependence of $\kc$ and $\kp$
of single crystals of {\ACoX} in zero field.
It is found that a peak appears at a low temperature below 10~K in
both $\kc$ of every {\ACoX} and $\kp$ of {\RCoC},
while a shoulder is observed around 60~K in only $\kc$ of every {\ACoX}.
Since {\ACoX} is an insulator,
the thermal conductivity is given by the sum of the contribution of
phonons and spin excitations.
The thermal conductivity due to phonons, $\kph$, is usually more isotropic than $\ksp$,
because the anisotropy of the correlation between atoms is generally much smaller
than that of the spin correlation in a low-dimensional spin system.
Moreover, the temperature dependence of $\kp$ is typical of $\kph$.
Therefore, it is concluded that the peak at a low temperature below 10~K is
due to $\kph$ and that the shoulder around 60~K is due to $\ksp$.
Similar temperature dependence of the thermal conductivity along spin chains
has been observed in several 1D AF Heisenberg spin systems
such as {\SCO} \cite{Sologubenko2000a,Sologubenko2001}
and {\SrCO}. \cite{Sologubenko2001}

In Fig.~\ref{fig:ACoX-kappa}, it is noticed that neither sharp dip is observed
at $\TNone$ nor at $\TNtwo$ for every {\ACoX},
while a sharp dip has been observed at the magnetic transition temperature
in every direction of thermal conductivity for {\BaCoVO}
\cite{Zhao2012,Kolland2013}
and several three-dimensional magnets.
\cite{Slack1958,Slack1961,Aring1967,Lewis1973}
The reason for neither sharp dip at $\TNone$ nor at $\TNtwo$ for {\ACoX} is
not clear at present, though several low-dimensional AF spin systems such as
{\SCO} \cite{Sologubenko2000a,Sologubenko2001}
and {\LaCuO} \cite{Sun2003}
have shown no sharp dip at their magnetic transition temperatures.

In Fig.~\ref{fig:ACoX-kappa}, it is also noticed that neither clear change is observed
at $\Tsone$ nor at $\Tstwo$ for {\RCoB}.
Since the structural phase transitions are caused by sliding motions of spin chains
along the $c$-axis,
it may be reasonable that $\kc$ is not affected by the transitions.

Here, we estimate the magnitude of $\ksp$.
Since $\ksp$ is given by the subtraction of $\kph$ from $\kc$ as follows:
\begin{equation}
	\ksp(T) = \kc(T) - \kph(T),
	\label{eq:ksp}
\end{equation}
the estimation of $\kph$ is necessary at first.
Based on the Debye model with the relaxation time approximation,
$\kph$ is given by
\begin{equation}
	\kph(T) = \frac{\Boltz}{2\pi^2 \vph}\left( \frac{\Boltz T}{\hbar} \right)^3 %
		\int_0^{\Debye/T} \D x \frac{x^4 e^x}{(e^x - 1)^2} \tph(x, T),
	\label{eq:Debye-kph}
\end{equation}
where $\Boltz$ is the Boltzmann constant, $\hbar$ the Planck constant,
$\Debye$ the Debye temperature, $\vph$ the velocity of phonons,
$\tph$ the relaxation time of the phonon with the angular frequency $\omega$,
and $x = \hbar\omega/\Boltz T$. \cite{Berman1976} 
The $\vph$ is calculated as
\begin{equation}
	\vph = \frac{\Boltz\Debye}{\hbar}\left( 6\pi^2 n \right)^{-1/3},
	\label{eq:vph}
\end{equation}
where $n$ is the number density of atoms.
The phonon scattering rate, $\tph^{-1}$, is given by the sum of
scattering rates in various scattering processes as follows:
\begin{equation}
	\tph^{-1} (\omega, T) = \frac{\vph}{L_{\rm b}} + P \omega^4 %
		+ D \omega + U \omega^2 T \exp(-\Debye / uT),
	\label{eq:Debye-tph}
\end{equation}
where $L_{\rm b}$, $P$, $D$, $U$ and $u$ are fitting parameters.
The first term represents the phonon scattering by boundaries.
The second term represents the phonon scattering by point defects.
The third term represents the phonon scattering by lattice distortions.
The fourth term represents the phonon-phonon scattering in the umklapp process.

\begin{figure}
	\includegraphics[width=0.95\textwidth]{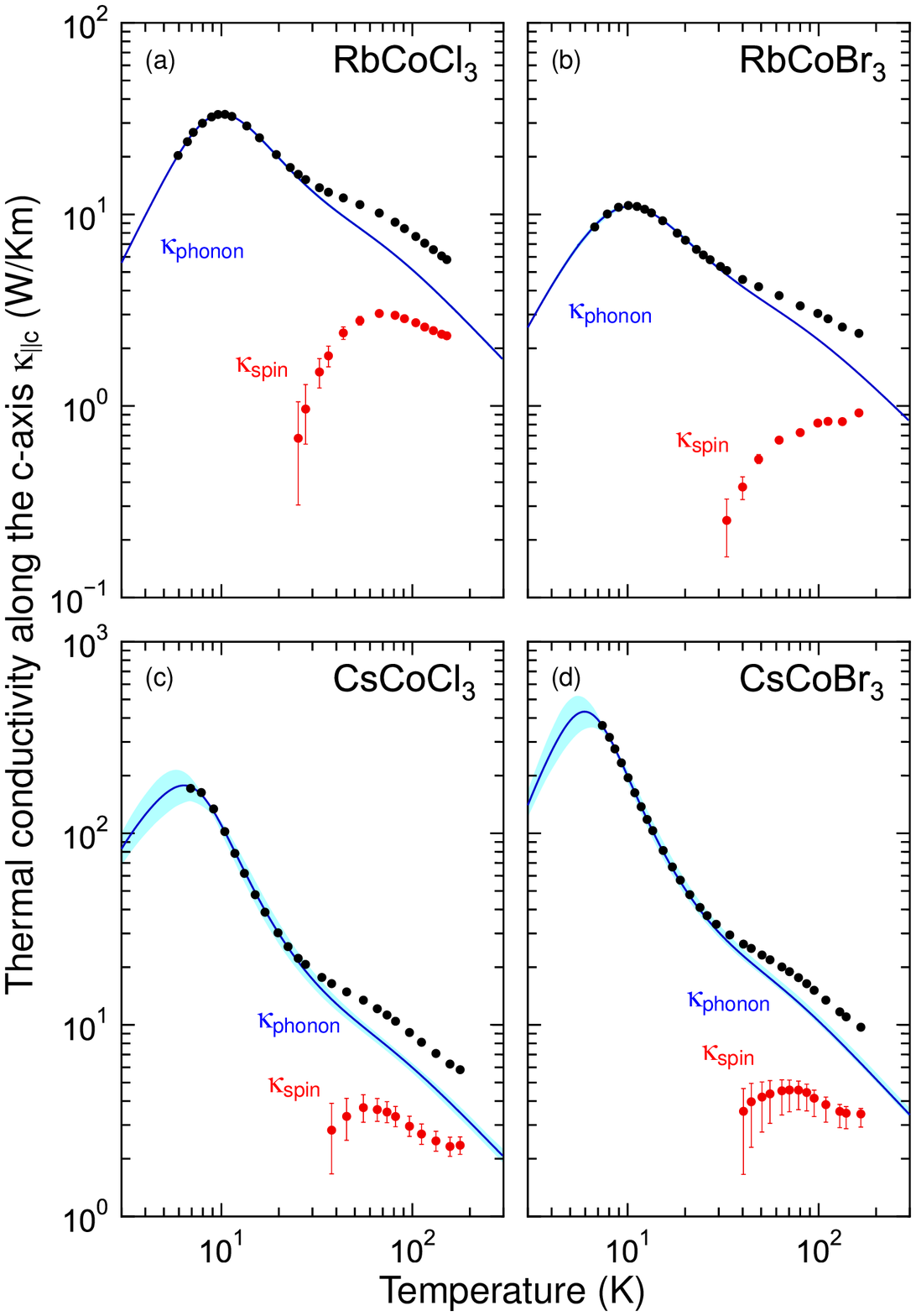}
	\caption{}\label{fig:ACoX-kappa-est}
\end{figure}

Using Eqs.~(\ref{eq:Debye-kph})--(\ref{eq:Debye-tph})
and putting $\Debye$ at 310~K estimated
from the M{\" o}ssbauer measurements, \cite{Bocquet1988}
the data of $\kp(T)$ in {\RCoC} are well fitted in a wide temperature-range
as shown by the black line in Fig.~\ref{fig:ACoX-kappa},
indicating that $\kp(T)$ is due to only $\kph(T)$.
The obtained parameters are listed in Table~\ref{tbl:kph-params}.
The estimation of $\kph(T)$ in $\kc(T)$ of {\ACoX} is performed
by the fit of the data of $\kc(T)$ at low temperatures below 25~K
with Eqs.~(\ref{eq:Debye-kph})--(\ref{eq:Debye-tph}),
because $\ksp(T)$ decreases markedly with decreasing temperature
at low temperatures of $\Boltz T \ll J$ in usual low-dimensional spin systems.
\cite{Kawamata2008}
In the fitting, values of $\Debye$ of {\RCoC} and {\CCoB} are put
at those estimated from the M{\" o}ssbauer measurements, \cite{Bocquet1988}
and those of {\RCoB} and {\CCoC} are put at the average value
of those of {\RCoC} and {\CCoB}, considering the weight of atoms.
The value of $L_{\rm b}$ is given by the distance between two terminals
of temperature on a single crystal.
The best-fit results of $\kph(T)$ are shown by blue solid lines
in Fig.~\ref{fig:ACoX-kappa-est}.
Shaded areas in Fig.~\ref{fig:ACoX-kappa-est} indicate errors of $\kph(T)$.
Values of parameters used for the best fit are listed in Table~\ref{tbl:kph-params}.
It is found that values of $P$ obtained in $\kc$ and $\kp$ of {\RCoC} are very different from each other, which is unreasonable. 
The difference of the $P$ value is caused by the analysis using the Debye model where the anisotropy of $\vph$ is not taken into account. 
That is, the anisotropy of $\vph$ appears to bring about the difference of the $P$ value between $\kc$ and $\kp$. 
Using Eq.~(\ref{eq:ksp}), $\ksp(T)$ is obtained as shown
by red circles in Fig.~\ref{fig:ACoX-kappa-est}.
It is found that $\ksp(T)$ exhibits the maximum at a high temperature above 50~K 
and that the maximum value of $\ksp(T)$ is about 3~W/Km in {\RCoC},
4~W/Km in {\CCoC}, {\CCoB},
and 0.9~W/Km in {\RCoB}.
Here, it is noted that $\ksp(T)$ is a little underestimated,
because $\ksp(T)$ is neglected at low temperatures below 25~K.
Values of $\ksp(T)$ in {\RCoB} are smaller than those of $\ksp(T)$ in the other $A$Co$X_3$. 
Since structural phase transitions occur at {$\Tsone$} and {$\Tstwo$} only in {\RCoB}, the lattice mismatch may be marked in {\RCoB} on account of the difference of the ionic radius between Rb$^+$ and Br$^-$. 
Therefore, the lattice mismatch may induce disorder of the lattice and suppress $\kph(T)$ and $\ksp(T)$. 

\begin{figure}
	\includegraphics[width=0.95\textwidth]{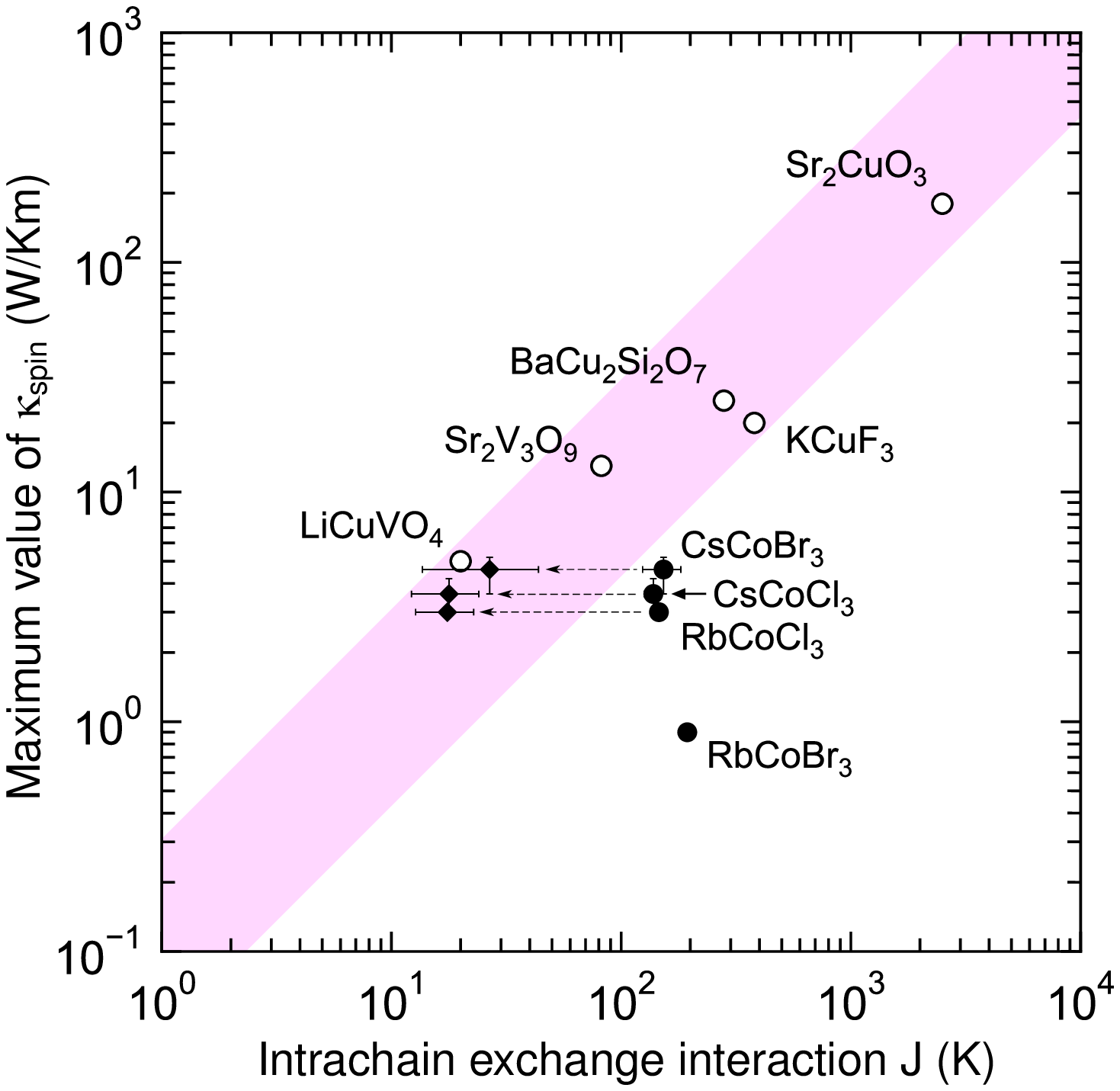}
	\caption{}\label{fig:ksp-J}
\end{figure}

Figure~\ref{fig:ksp-J} displays the maximum values of $\ksp$ in {\ACoX}
and several 1D AF Heisenberg spin systems
\cite{Miike1975a,Sologubenko2003a,Hlubek2012,Parfeneva2004,Uesaka2010}
plotted as a function of $J$.
It is found that the maximum values of $\ksp$ in 1D AF Heisenberg spin systems
tend to be proportional to $J$.
This is simply understood to be due to the velocity of spin excitations, $\vsp$,
being proportional to $J$
in spite of the specific heat, $\Csp$, being inversely proportional to $J$
at low temperatures in 1D AF Heisenberg spin systems,
\cite{DesCloizeaux1962,Takahashi1973}
because $\ksp$ is proportional to $\Csp\vsp^2$.
It is found that the maximum values of $\ksp$ in {\ACoX} are located
a little below the shaded zone indicating the rough proportionality
between the maximum value of $\ksp$ and $J$.
The bandwidth of the magnetic dispersion in an 1D Heisenberg spin system
is proportional to $J$,
so that $\vsp$ is proportional to $J$ at low temperatures.
In an 1D AF quasi-Ising spin system, on the other hand,
the bandwidth, namely, $\vsp$ is expected to be proportional to
$\varepsilon J$, according to Eq.~(\ref{eq:H2}).
\cite{comment01}
Using values of $\varepsilon J$ instead of those of $J$ in Fig.~\ref{fig:ksp-J},
therefore, the maximum values of $\ksp$ in {\ACoX} are found to shift just into the shaded zone.
Accordingly, this result strongly suggests that a little Heisenberg-like character
with $\varepsilon \neq 0$ brings about a significant bandwidth,
leading to the appearance of $\ksp$ even in an 1D AF Ising-like spin system.
Here, it is noted that $\kc$ of {\RCoC} little changes by the application
of magnetic field of 14~T along the $c$-axis in a wide temperature-range
of 6--140~K, as shown in Fig.~\ref{fig:ACoX-kappa}.
That is, $\ksp$ appears not to be affected by the magnetic field of 14~T.
The magnetic field of 14~T may be too small to affect $\ksp$ of {\RCoC} being
markedly observed at high temperatures around 70~K.
Furthermore, it is not clear why no $\ksp$ has been observed in the quasi-1D
AF spin system {\BaCoVO}
in spite of the sizable value of $\varepsilon \sim 0.5$.
To be more conclusive, therefore, thermal conductivity measurements
in the other 1D AF Ising-like spin systems are necessary.

\section{Summary}

Thermal conductivity measurements have been carried out for single crystals
of the 1D AF Ising-like spin system {\ACoX} {\AXset}.
It has been found that $\kp$ perpendicular to spin chains shows
an only peak at a low temperature below 10~K
and that $\kc$ parallel to spin chains shows a shoulder around 60~K
besides a peak at a low temperature below 10~K.
It has been concluded that the peak is due to $\kph$
and that the shoulder around 60~K is due to $\ksp$.
To our knowledge, this is the first observation of $\ksp$
in 1D AF Ising-like spin systems.
It has been found that the magnitude of $\ksp$ in {\ACoX}
estimated by the subtraction of $\kph$ from $\kc$ is a little smaller
than those in 1D AF Heisenberg spin systems
and that the magnitude of $\ksp$ tends to be proportional to $\varepsilon J$
universally in both {\ACoX} and 1D AF Heisenberg spin systems.
Accordingly, it has been concluded that a little Heisenberg-like character
with $\varepsilon \neq 0$ brings about a significant bandwidth of spin excitations,
leading to the appearance of $\ksp$ even in an 1D AF Ising-like spin system.
To be more conclusive, thermal conductivity measurements
in the other 1D AF Ising-like spin systems are necessary.

\begin{acknowledgment}

The thermal conductivity measurements in magnetic fields were performed
at the High Field Laboratory for Superconducting Materials,
Institute for Materials Research, Tohoku University.

\end{acknowledgment}

\newpage

\section*{Figure captions}

\begin{description}
\item{Fig.~1.}
		(Color online) Temperature dependence of the thermal conductivity along the $c$-axis,
		$\kc$, in {\ACoX} {\AXset} and
		the thermal conductivity in the $c$-plane, $\kp$, in {\RCoC} in zero field.
		The black line is the best-fit result of $\kp$ in {\RCoC} using		
		the thermal conductivity due to phonons given by
		Eqs.~(\ref{eq:Debye-kph})--(\ref{eq:Debye-tph})
		with parameters listed in Table~\ref{tbl:kph-params}.
		Red circles show {$\kc$} of {\RCoC} in a magnetic field of 14~T
		applied along the $c$-axis.
\item{Fig.~2.}
		(Color online) Temperature dependence of the thermal conductivity along the $c$-axis, $\kc$,
		in (a)~{\RCoC}, (b)~{\RCoB}, (c)~{\CCoC}, (d)~{\CCoB}.
		Blue solid lines show {$\kph$} estimated
		using Eqs.~(\ref{eq:Debye-kph})--(\ref{eq:Debye-tph}).
		Shaded areas demonstrate errors of {$\kph$}.
		Red circles show {$\ksp$} obtained by subtracting {$\kph$} from {$\kc$}.
\item{Fig.~3.}
		(Color online) Dependence on the intrachain interaction, $J$, of the maximum value of $\ksp$
		in the 1D AF Ising-like spin system {\ACoX} {\AXset} (closed circles)
		and in 1D AF Heisenberg spin systems (open circles).
		\cite{Miike1975a,Sologubenko2003a,Hlubek2012,Parfeneva2004,Uesaka2010}
		Diamonds show dependence on $\varepsilon J$ of the maximum value
		of $\ksp$ in {\ACoX}.
		The shaded zone indicates the rough proportionality
		between the maximum value of $\ksp$ and $J$.
\end{description}

\newpage

\begin{table}
	\caption{
		Magnetic parameters
		(the intrachain exchange interaction, $J$,
		the anisotropy parameter of spin, $\varepsilon$,
		the ratio of the interchain exchange interaction, $J'$, to $J$,
		the magnetic phase transition temperatures,
		$\TNone$ and $\TNtwo$)
		and structural parameters
		(the structural phase transition temperatures, $\Tsone$ and $\Tstwo$)
		of {\ACoX} {\AXset}.
	}
	\label{tbl:ACoX-param}
\begin{center}
\begin{tabular}{cccccccc} \hline
 	& {$J$} (K) & {$\varepsilon$} & {$J'/J$}
	& {$\TNone$} (K) & {$\TNtwo$} (K) & {$\Tsone$} (K) & {$\Tstwo$} (K) \\ \hline\hline
	{\RCoC}\fnm[1]
	& 140--152 & 0.091--0.15 & 0.02--0.11 & 28 & 11 & - & - \\
	{\RCoB}\fnm[2]
	& 194 & - & 0.025 & 37 & 31 & 90 & 37 \\	
	{\CCoC}\fnm[3]
	& 126--150 & 0.097--0.16 & 0.013--0.12 & 21 & 9 & - & -  \\
	{\CCoB}\fnm[4]
	& 124--182 & 0.11--0.24 & 0.019--0.13 & 28 & 10 & - & - \\ \hline
\end{tabular}
	\vspace{1mm}
	{\footnotesize
	\fnm[1]~{Ref.~\onlinecite{Lockwood1983,Matsubara1991a,Hori1990}.}\:\:\:\:
	\fnm[2]~{Ref.~\onlinecite{Nishiwaki2008,Yamanaka2002a}.}\:\:\:\:
	\fnm[3]~{Ref.~\onlinecite{Lehmann1981,Matsubara1991a,Yoshizawa1981,Nagler1983,
	  Matsubara1991,Hori1990,Shiba2003,Yoshizawa1979}.}\:\:\:\:
	\fnm[4]~{Ref.~\onlinecite{Lehmann1981,Matsubara1991a,Nagler1983,Hori1990,Yelon1975}.}
	}
\end{center}
\end{table}

\begin{table}
\caption{
	Parameters used for the fit of the temperature dependences of the thermal conductivity,
	$\kc$ and $\kp$, in {\ACoX} with Eqs.~(\ref{eq:Debye-kph})--(\ref{eq:Debye-tph}).
}
\label{tbl:kph-params}
 \begin{tabular}{cccccccc} \hline
				& $\kappa$ & $\Debye$ (K) & $L_{\rm b}$ (mm) & $P$ ($10^{-43}$ s$^3$) %
				& $D$ ($10^{-6}$) & $U$ ($10^{-18}$ sK$^{-1}$) & $u$ \\ \hline\hline
	\multirow{2}{*}{\RCoC}
				& $\kp$ & 310 & 1.48 & 52.5 {\scrs $\pm$ 1.9} & 36.1 {\scrs $\pm$ 1.0}
				& 39.8 {\scrs $\pm$ 0.4} & 7.66 {\scrs $\pm$ 0.05} \\
				& $\kc$ & 310 & 2.03 & 0.82 {\scrs $\pm$ 0.35} & 43.3 {\scrs $\pm$ 1.3}
				& 19.0 {\scrs $\pm$ 0.4} & 9.8 {\scrs $\pm$ 0.2} \\ \hline
	\RCoB	& $\kc$ & 300 & 1.26 & 16.9 {\scrs $\pm$ 3.8} & 86.0 {\scrs $\pm$ 5.7}
				& 35.0 {\scrs $\pm$ 1.0} & 10.0 {\scrs $\pm$ 0.4} \\ \hline
	\CCoC	& $\kc$ & 300 & 1.65 & 3.6 {\scrs $\pm$ 5.6} & 0 {\scrs $\pm$ 2.5}
				& 15.5 {\scrs $\pm$ 1.7} & 8.3 {\scrs $\pm$ 0.8 }\\ \hline
	\CCoB	& $\kc$ & 290 & 1.54 & 0.24 {\scrs $\pm$ 1.40} & 0.08 {\scrs $\pm$ 2.69}
				& 9.23 {\scrs $\pm$ 0.69} & 8.82 {\scrs $\pm$ 0.50} \\ \hline
  \end{tabular}
\end{table}

\newpage

\newpage

\begin{thebibliography}{10}

\bibitem{Miike1975a}
H.~Miike and K.~Hirakawa, J.~Phys. Soc. Jpn. {\bfseries 38},  1279 (1975).

\bibitem{Kudo1999}
K.~Kudo, S.~Ishikawa, T.~Noji, T.~Adachi, Y.~Koike, K.~Maki, S.~Tsuji, and
  K.~Kumagai, J. Low Temp. Phys. {\bfseries 117},  1689 (1999).

\bibitem{Sologubenko2000}
A.~V. Sologubenko, K.~Giann{\` o}, H.~R. Ott, U.~Ammerahl, and A.~Revcolevschi,
  Phys. Rev. Lett. {\bfseries 84},  2714 (2000).

\bibitem{Sologubenko2000b}
A.~V. Sologubenko, K.~Giann{\` o}, H.~R. Ott, U.~Ammerahl, A.~Revcolevschi,
  D.~F. Brewer, and A.~L. Thomson, Physica B {\bfseries 284-288},  1595 (2000).

\bibitem{Sologubenko2000a}
A.~V. Sologubenko, E.~Felder, K.~Giann{\` o}, H.~R. Ott, A.~Vietkine, and
  A.~Revcolevschi, Phys. Rev. B {\bfseries 62},  R6108 (2000).

\bibitem{Sologubenko2001}
A.~V. Sologubenko, K.~Giann{\` o}, H.~R. Ott, A.~Vietkine, and A.~Revcolevschi,
  Phys. Rev. B {\bfseries 64},  054412 (2001).

\bibitem{Kudo2001}
K.~Kudo, S.~Ishikawa, T.~Noji, T.~Adachi, Y.~Koike, K.~Maki, S.~Tsuji, and
  K.~Kumagai, J.~Phys. Soc. Jpn. {\bfseries 70},  437 (2001).

\bibitem{Kudo2001a}
K.~Kudo, Y.~Koike, K.~Maki, S.~Tsuji, and K.~Kumagai, J.~Phys. Chem. Solids
  {\bfseries 62},  361 (2001).

\bibitem{Hess2001}
C.~Hess, C.~Baumann, U.~Ammerahl, B.~B{\" u}chner, F.~Heidrich-Meisner,
  W.~Brenig, and A.~Revcolevschi, Phys. Rev. B {\bfseries 64},  184305 (2001).

\bibitem{Sologubenko2003a}
A.~V. Sologubenko, H.~R. Ott, G.~Dhalenne, and A.~Revcolevschi, Europhys. Lett.
  {\bfseries 62},  540 (2003).

\bibitem{Hess2004a}
C.~Hess, H.~ElHaes, B.~B{\" u}chner, U.~Ammerahl, M.~H{\" u}cker, and
  A.~Revcolevschi, Phys. Rev. Lett. {\bfseries 93},  027005 (2004).

\bibitem{Hess2006}
C.~Hess, P.~Ribeiro, B.~B{\" u}chner, H.~ElHaes, G.~Roth, U.~Ammerahl, and
  A.~Revcolevschi, Phys. Rev. B {\bfseries 73},  104407 (2006).

\bibitem{Kawamata2008}
T.~Kawamata, N.~Takahashi, T.~Adachi, T.~Noji, K.~Kudo, N.~Kobayashi, and
  Y.~Koike, J.~Phys. Soc. Jpn. {\bfseries 77},  034607 (2008).

\bibitem{Kawamata2010}
T.~Kawamata, N.~Kaneko, M.~Uesaka, M.~Sato, and Y.~Koike, J.~Phys.: Conf. Ser.
  {\bfseries 200},  022023 (2010).

\bibitem{N.Hlubek2010}
N.~Hlubek, Dr. Thesis, Technische Universtit{\" a}t Dresden (2010).

\bibitem{Hlubek2010}
N.~Hlubek, P.~Ribeiro, R.~Saint-Martin, A.~Revcolevschi, G.~Roth, G.~Behr,
  B.~B{\" u}chner, and C.~Hess, Phys. Rev. B {\bfseries 81},  020405 (2010).

\bibitem{Hlubek2012}
N.~Hlubek, X.~Zotos, S.~Singh, R.~Saint-Martin, A.~Revcolevschi, B.~B{\"
  u}chner, and C.~Hess, J. Stat. Mech.  P03006 (2012).

\bibitem{Nakamura1991}
Y.~Nakamura, S.~Uchida, T.~Kimura, N.~Motohira, K.~Kishio, K.~Kitazawa,
  T.~Arima, and Y.~Tokura, Physica C {\bfseries 185-189},  1409 (1991).

\bibitem{Takenaka1997}
K.~Takenaka, Y.~Fukuzumi, K.~Mizuhashi, S.~Uchida, H.~Asaoka, and H.~Takei,
  Phys. Rev. B {\bfseries 56},  5654 (1997).

\bibitem{Sun2003}
X.~F. Sun, J.~Takeya, S.~Komiya, and Y.~Ando, Phys. Rev. B {\bfseries 67},
  104503 (2003).

\bibitem{Yan2003}
J.-Q. Yan, J.-S. Zhou, and J.~B. Goodenough, Phys. Rev. B {\bfseries 68},
  104520 (2003).

\bibitem{Hess2003a}
C.~Hess, B.~B{\" u}chner, U.~Ammerahl, L.~Colonescu, F.~Heidrich-Meisner,
  W.~Brenig, and A.~Revcolevschi, Phys. Rev. Lett. {\bfseries 90},  197002
  (2003).

\bibitem{Hess2003}
C.~Hess, B.~B{\" u}chner, U.~Ammerahl, and A.~Revcolevschi, Phys. Rev. B
  {\bfseries 68},  184517 (2003).

\bibitem{Hofmann2003}
M.~Hofmann, T.~Lorenz, K.~Berggold, M.~Gr{\" u}ninger, A.~Freimuth, G.~S.
  Uhrig, and E.~Br{\" u}ck, Phys. Rev. B {\bfseries 67},  184502 (2003).

\bibitem{Hess2004}
C.~Hess and B.~B{\" u}chner, Eur. Phys. J. B {\bfseries 38},  37 (2004).

\bibitem{Berggold2006}
K.~Berggold, T.~Lorenz, J.~Baier, M.~Kriener, D.~Senff, H.~Roth, A.~Severing,
  H.~Hartmann, A.~Freimuth, S.~Barilo, and F.~Nakamura, Phys. Rev. B {\bfseries
  73},  104430 (2006).

\bibitem{Kudo2004}
K.~Kudo, Y.~Koike, S.~Kurogi, T.~Noji, T.~Nishizaki, and N.~Kobayashi, J. Magn.
  Magn. Mater. {\bfseries 272-276},  94 (2004).

\bibitem{Choi2004}
J.-H. Choi, T.~C. Messina, J.~Yan, G.~I. Drandova, and J.~T. Markert, J. Magn.
  Magn. Mater. {\bfseries 272-276},  970 (2004).

\bibitem{Huber1969}
D.~L. Huber and J.~S. Semura, Phys. Rev. {\bfseries 182},  602 (1969).

\bibitem{Zhao2012}
Z.~Y. Zhao, X.~G. Liu, Z.~Z. He, X.~M. Wang, C.~Fan, W.~P. Ke, Q.~J. Li, L.~M.
  Chen, X.~Zhao, and X.~F. Sun, Phys. Rev. B {\bfseries 85},  134412 (2012).

\bibitem{Kolland2013}
G.~Kolland, Dr. Thesis, Universit{\" a}t zu K{\" o}ln (2013).

\bibitem{Lehmann1981}
W.~P. Lehmann, W.~Breitling, and R.~Weber, J.~Phys. C: Solid State Phys.
  {\bfseries 14},  4655 (1981).

\bibitem{Lockwood1983}
D.~J. Lockwood, I.~W. Johnstone, H.~J. Labbe, and B.~Briat, J.~Phys. C: Solid
  State Phys. {\bfseries 16},  6451 (1983).

\bibitem{Matsubara1991a}
F.~Matsubara, S.~Inawashiro, and H.~Ohhara, J.~Phys.: Condens. Matter
  {\bfseries 3},  1815 (1991).

\bibitem{Yoshizawa1981}
H.~Yoshizawa, K.~Hirakawa, S.~K. Satija, and G.~Shirane, Phys. Rev. B
  {\bfseries 23},  2298 (1981).

\bibitem{Nagler1983}
S.~E. Nagler, W.~J.~L. Buyers, R.~L. Armstrong, and B.~Briat, Phys. Rev. B
  {\bfseries 27},  1784 (1983).

\bibitem{Matsubara1991}
F.~Matsubara and S.~Inawashiro, Phys. Rev. B {\bfseries 43},  796 (1991).

\bibitem{Hori1990}
H.~Hori, H.~Mikami, M.~Date, and K.~Amaya, Physica B {\bfseries 165-166},  237
  (1990).

\bibitem{Shiba2003}
H.~Shiba, Y.~Ueda, K.~Okunishi, S.~Kimura, and K.~Kindo, J.~Phys. Soc. Jpn.
  {\bfseries 72},  2326 (2003).

\bibitem{Yoshizawa1979}
H.~Yoshizawa and K.~Hirakawa, J.~Phys. Soc. Jpn. {\bfseries 46},  448 (1979).

\bibitem{Yelon1975}
W.~B. Yelon, D.~E. Cox, and M.~Eibsch{\" u}tz, Phys. Rev. B {\bfseries 12},
  5007 (1975).

\bibitem{Nishiwaki2008}
Y.~Nishiwaki, T.~Nakamura, A.~Oosawa, K.~Kakurai, N.~Todoroki, N.~Igawa,
  Y.~Ishii, and T.~Kato, J.~Phys. Soc. Jpn. {\bfseries 77},  104703 (2008).

\bibitem{Yamanaka2002a}
K.~Yamanaka, Y.~Nishiwaki, K.~Iio, T.~Kato, T.~Mitsui, T.~Tojo, and T.~Atake,
  J. Therm. Anal. Calorim. {\bfseries 70},  371 (2002).

\bibitem{Slack1958}
G.~A. Slack and R.~Newman, Phys. Rev. Lett. {\bfseries 1},  359 (1958).

\bibitem{Slack1961}
G.~A. Slack, Phys. Rev. {\bfseries 122},  1451 (1961).

\bibitem{Aring1967}
K.~Aring and A.~J. Sievers, J. Appl. Phys. {\bfseries 38},  1496 (1967).

\bibitem{Lewis1973}
F.~B. Lewis and N.~H. Saunders, J.~Phys. C: Solid State Phys. {\bfseries 6},
  2525 (1973).

\bibitem{Berman1976}
R.~Berman, {\em Thermal Conduction in Solids} (Clarendon Press, Oxford, U. K.,
  1976).

\bibitem{Bocquet1988}
S.~Bocquet, J.~B. Ward, and V.~H. McCann, J.~Phys. C: Solid State Phys.
  {\bfseries 21},  367 (1988).

\bibitem{Parfeneva2004}
L.~S. Parfen'eva, I.~A. Smirnov, H.~Misiorek, J.~Mucha, A.~Jezowski, A.~V.
  Prokof'ev, and W.~Assmus, Phys. Solid State {\bfseries 46},  357 (2004).

\bibitem{Uesaka2010}
M.~Uesaka, T.~Kawamata, N.~Kaneko, M.~Sato, K.~Kudo, N.~Kobayashi, and
  Y.~Koike, J.~Phys.: Conf. Ser. {\bfseries 200},  022068 (2010).

\bibitem{DesCloizeaux1962}
J.~des Cloizeaux and J.~J. Pearson, Phys. Rev. {\bfseries 128},  2131 (1962).

\bibitem{Takahashi1973}
M.~Takahashi, Prog. Theor. Phys. {\bfseries 50},  1519 (1973).

\bibitem{comment01}
Theoretically, two kinds of spin excitations are known to exist in an 1D AF
  Ising-like spin system with $\varepsilon \neq 0$: one is the Villain mode
  with the bandwidth of $2 \varepsilon J$, \cite{Villain1975} and the other is
  the spin-wave mode with the bandwidth of $4\varepsilon J$
  \cite{Ishimura1980}.

\bibitem{Villain1975}
J.~Villain, Physica B+C {\bfseries 79},  1 (1975).

\bibitem{Ishimura1980}
N.~Ishimura and H.~Shiba, Prog. Theor. Phys. {\bfseries 63},  743 (1980).

\end{thebibliography}
\end{document}